\documentclass[letterpage, 11pt]{article}
\usepackage{comment}
\usepackage{lipsum}
\usepackage[pdfborder={0 0 1}]{hyperref}
\usepackage{rotating} 
\usepackage[margin=1in]{geometry}
\usepackage{amsmath,amssymb,wasysym,algorithm,algorithmic}
\usepackage{graphicx}
\usepackage{float}
\usepackage{authblk}
\usepackage{xspace}
\usepackage{caption}
\usepackage{booktabs}
\usepackage{fancyhdr}
\usepackage[section]{placeins}
\usepackage[T1]{fontenc}

\title{Modeling the Impact of Exposed Cases in a Hantavirus Outbreak on a Cruise Ship}
\date{}
\author[1]{Jiaming Cui}

\affil[1]{Department of Computer Science, Virginia Tech}

\begin{document}
\maketitle

\begin{abstract}
The emergence of a hantavirus variant aboard a commercial cruise ship presents a significant public health concern. This study develops a discrete-time stochastic Susceptible-Exposed-Infectious-Recovered-Dead model to estimate transmission dynamics, hidden exposed infections, and outbreak risk among passengers and crew. Epidemiological parameters and latent disease states were inferred using an Ensemble Adjustment Kalman Filter calibrated to reported case data from WHO and ECDC situation reports. The estimated basic reproduction number was 2.76, with a 95\% confidence interval of 2.52-2.99, indicating substantial potential for sustained onboard transmission before strict quarantine measures. Simulations further suggest that several exposed individuals may remain unidentified during the early outbreak phase, creating a hidden reservoir that symptom-based surveillance alone may fail to detect. These findings highlight the importance of rapid surveillance, widespread testing, targeted quarantine, and active monitoring of exposed individuals in confined travel settings. The proposed modeling framework can support timely outbreak assessment and intervention planning for infectious-disease events in similarly dense and spatially constrained populations.
\end{abstract}

\section{Introduction}

The recent emergence and rapid spread of a hantavirus variant aboard a commercial cruise ship present a distinct and urgent public health challenge~\cite{WHO,ECDC}. Cruise ships are densely populated, spatially constrained environments in which passengers and crew share dining areas, recreational facilities, corridors, cabins, and service spaces over prolonged periods~\cite{kak2007infections,pavli2016respiratory,hu2013scaling,zhang2016contact}. Although confirmed symptomatic cases can be isolated rapidly once identified, transmission may continue if a substantial number of exposed individuals remain undetected~\cite{rocklov2020covid,batista2020minimizing,emery2020contribution}. This concern is especially important for pathogens with a long interval between exposure and symptom onset, as infected individuals may remain outside the case-identification system for an extended period before clinical signs appear, and may further spread the disease after disembarkation into larger communities~\cite{pavli2016respiratory,manigold2014human,young2000incubation}.

Despite the central importance of exposed individuals, little is known about their number, timing, and contribution to outbreak growth in this setting. Confirmed case counts provide information only after infection has progressed to detectable illness or diagnostic confirmation, whereas the exposed population remains largely hidden~\cite{mizumoto2020estimating,zhang2020estimation}. This gap limits the ability to determine whether observed cases represent isolated events or the visible portion of a larger, delayed epidemic process. In a finite and highly connected population such as a cruise ship, even a modest underestimation of the exposed pool may lead to substantial errors in projected outbreak size and in the evaluation of quarantine policies.

Therefore, this paper develops a mathematical modeling framework to estimate the hidden burden and transmission consequences of exposed infections aboard the vessel~\cite{vial2006incubation,diekmann1990definition,van2002reproduction,hethcote2000mathematics}. We propose a customized SEIRD model~\cite{chopra2022differentiable} that divides the population into susceptible, exposed, infectious, recovered, and deceased compartments, with particular emphasis on the transition from exposure to symptomatic infection. This framework allows us to represent the delay between infection and detection and to evaluate how unobserved exposed individuals may sustain transmission before they are recognized as confirmed cases.

By simulating outbreak trajectories under different assumptions, we estimate that several exposed infections may remain unidentified during the early phase of the outbreak. We further estimate the basic reproduction number, R0, to be 2.76, with a 95\% confidence interval of 2.52-2.99, which is similar to previous results on other Hantavirus outbreak~\cite{martinez2020super}. Because R0 exceeds 1, this finding suggests the potential for sustained transmission within the cruise-ship setting. This estimate should not be directly generalized to land-based communities, where population density, contact structure, and exposure patterns may differ substantially. Nevertheless, an R0 greater than 1 highlights the need for rapid surveillance, targeted quarantine, and further assessment of exposed individuals. Overall, this study provides an initial modeling framework for hantavirus outbreak assessment in spatially constrained populations and may support timely public health decision-making in similar high-density settings.
\section{Methods}

\subsection{Stochastic SEIRD Transmission Model}

\begin{figure}
    \centering
    \includegraphics[width=0.4\linewidth]{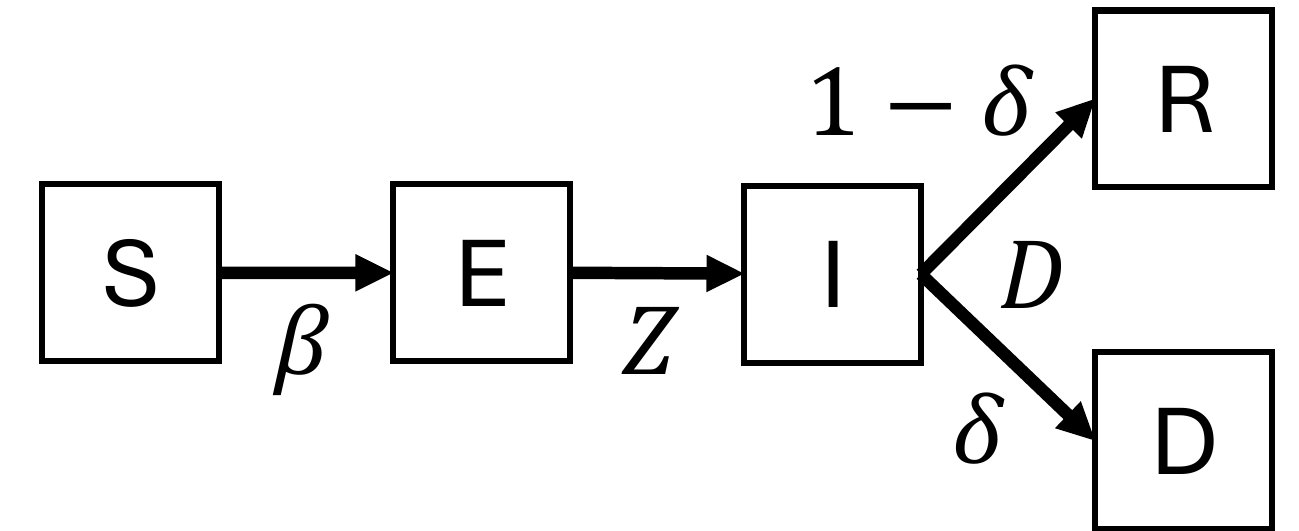}
    \caption{The diagram of the SEIRD model}
    \label{fig:diagram}
\end{figure}

To capture the transmission dynamics of the Hantavirus variant within the closed environment of the cruise ship, we formulated a discrete-time, stochastic Susceptible-Exposed-Infectious-Recovered-Dead (SEIRD) compartmental model as shown in Figure~\ref{fig:diagram}. We assume that transmission occurs exclusively through contact with individuals in the infectious compartment, denoted by $I$. The exposed compartment $E$ represents individuals who have been infected but have not yet developed symptoms, whereas the infectious compartment $I$ represents symptomatic and documented cases.

Let $S_t$, $E_t$, $I_t$, $R_t$, and $D_t$ denote the numbers of susceptible, exposed, infectious, recovered, and deceased individuals on day $t$, respectively, and let $N_t = S_t + E_t + I_t + R_t + D_t$. The corresponding mean-field discrete-time dynamics are given by
\begin{align*}
\Delta S_t &= -\beta \frac{S_t I_t}{N_t}, \\
\Delta E_t &= \beta \frac{S_t I_t}{N_t} - \frac{E_t}{Z}, \\
\Delta I_t &= \frac{E_t}{Z} - \frac{I_t}{D}, \\
\Delta R_t &= (1-\delta)\frac{I_t}{D}, \\
\Delta D_t &= \delta \frac{I_t}{D},
\end{align*}
where $\Delta X_t = X_{t+1}-X_t$, $\beta$ is the transmission rate, $Z$ is the average duration of the exposed period, $D$ is the average duration of the infectious period, and $\delta$ is the case fatality rate. To account for stochasticity in transmission and disease progression within a relatively small, confined population, the daily transition counts are modeled as Poisson random variables. Specifically,
\begin{enumerate}
    \item \textbf{New exposures} $(U_{1,t})$: New infections are driven by the transmission rate $\beta$ and contact between susceptible and infectious individuals: $U_{1,t} \sim \text{Poisson}\left(\beta \frac{S_t I_t}{N_t}\right)$.
    \item \textbf{Symptom onset / documentation} $(U_{2,t})$: Progression from the exposed compartment to the infectious compartment is governed by the average exposed period $Z$: $U_{2,t} \sim \text{Poisson}\left(\frac{E_t}{Z}\right)$.
    \item \textbf{Recoveries and fatalities} $(U_{3,t}, U_{4,t})$: Removal from the infectious compartment is governed by the average infectious period $D$, with outcomes determined by the case fatality rate $\delta$: $U_{3,t} \sim \text{Poisson}\left((1-\delta)\frac{I_t}{D}\right),
    \qquad
    U_{4,t} \sim \text{Poisson}\left(\delta \frac{I_t}{D}\right)$.
\end{enumerate}

The stochastic SEIRD model can therefore be written as the following set of daily update equations:
\begin{align*}
S_{t+1} &= S_t - U_{1,t}, \\
E_{t+1} &= E_t + U_{1,t} - U_{2,t}, \\
I_{t+1} &= I_t + U_{2,t} - U_{3,t} - U_{4,t}, \\
R_{t+1} &= R_t + U_{3,t}, \\
D_{t+1} &= D_t + U_{4,t}.
\end{align*}

\subsection{Parameter Inference}

Epidemiological data for the cruise ship outbreak were obtained from official World Health Organization (WHO) situation reports. The dataset includes daily counts of confirmed cases, recoveries, and fatalities. Parameter values and admissible ranges were informed by previous hantavirus modeling studies, as well as publicly available WHO and European Centre for Disease Prevention and Control (ECDC) reports. The parameter ranges and estimated values are summarized in Table~\ref{tab:parameter}.

\begin{table}[h]
    \centering
    \caption{List of parameters}
    \begin{tabular}{c|c|c}
        \toprule
        Parameter & range & estimated value (95\% CI) \\
        \midrule
        Transmission rate ($\beta$, days$^{-1}$) & [0.1,0.5]~\cite{martinez2020super} & 0.23 (0.22,0.25) \\
        Latency period ($Z$, days) & [6,12]~\cite{vial2006incubation} & 9.12 (8.76,9.48) \\
        Infectious period ($D$, days) & [7,14]~\cite{WHO,ECDC} & 11.52 (11.06,11.97) \\
        factality rate ($\delta$) & [0.3,0.5]~\cite{lazaro2007clusters} & 0.36 (0.35,0.38) \\
        \bottomrule
    \end{tabular}
    \label{tab:parameter}
\end{table}

For model calibration, we used the Ensemble Adjustment Kalman Filter (EAKF) to infer latent epidemiological states and estimate key model parameters. The EAKF is a recursive Bayesian data assimilation method that is well suited for nonlinear infectious disease models. By sequentially assimilating daily reported incidence data from WHO/ECDC reports into the stochastic SEIRD framework, the EAKF dynamically estimates unobserved state variables, such as the true number of exposed individuals, while simultaneously updating the posterior distributions of epidemiological parameters. Specifically, we fitted the cumulative number of transitions from the exposed compartment $E$ to the infectious compartment $I$ to the officially reported cumulative case counts. For the simulations, the EAKF was run for 10 iterations using an ensemble size of 300, following previous studies~\cite{li2020substantial,pei2020differential}.
\section{Results}

\subsection{Model fitting}

\begin{figure}[h]
    \centering
    \includegraphics[width=0.6\linewidth]{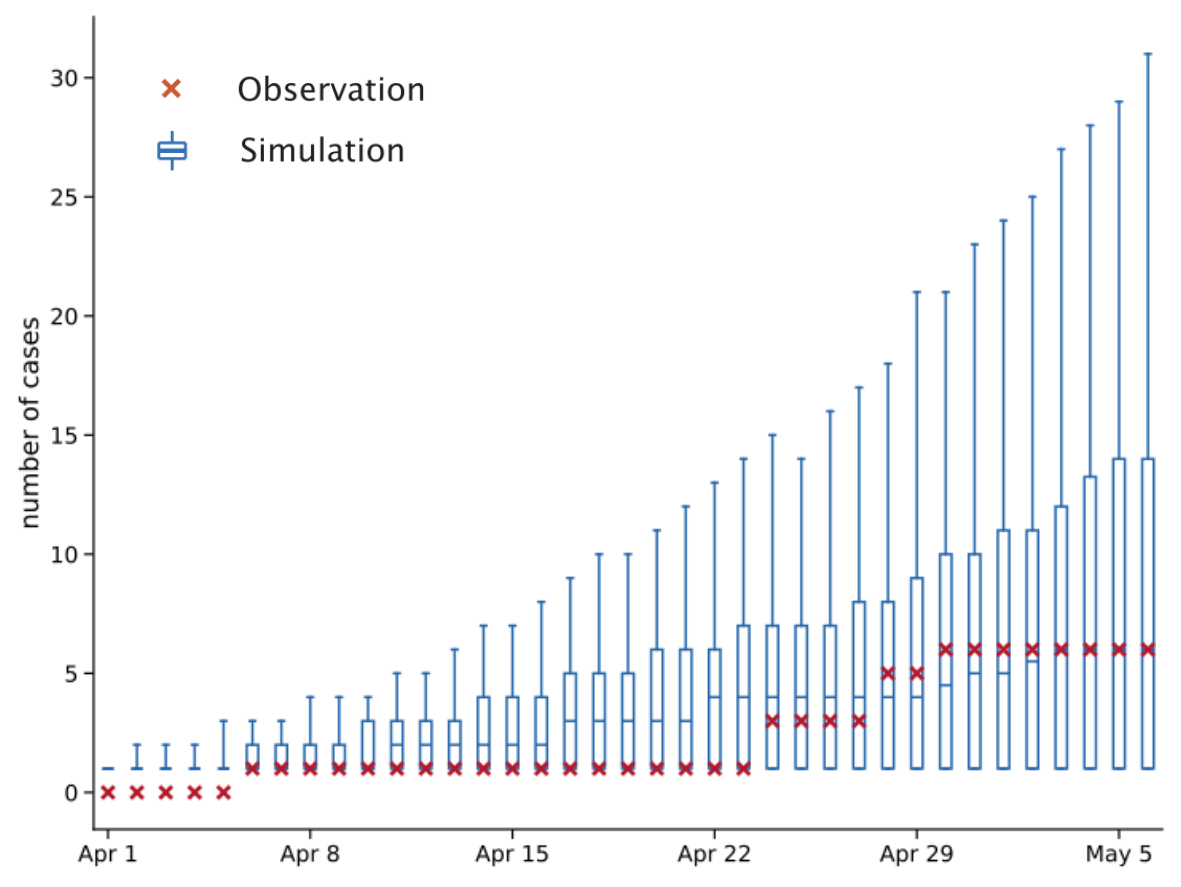}
    \caption{Simulated number of identified cases on board. The blue box-and-whisker plots indicate the median, interquartile range, and 95\% simulation intervals.}
    \label{fig1}
\end{figure}

Figure~\ref{fig1} compares the cumulative reported cases with the simulated trajectories generated by our model over the course of the outbreak, from April 1 to May 7. As shown in the figure, the distribution of stochastic simulations closely captures the observed epidemiological data, indicated by the red crosses. In particular, the model reproduces both the slow accumulation of documented cases in early April and the subsequent stepwise increases observed in late April. The strong agreement between the reported case counts and the simulated distributions suggests that the model-inference framework can reliably reconstruct the underlying transmission dynamics and capture the observed variability within the enclosed cruise-ship environment. The estimated parameter values are listed in Table~\ref{tab:parameter}.

\subsection{Reproductive number estimation}

\begin{figure}[h]
    \centering
    \includegraphics[width=\linewidth]{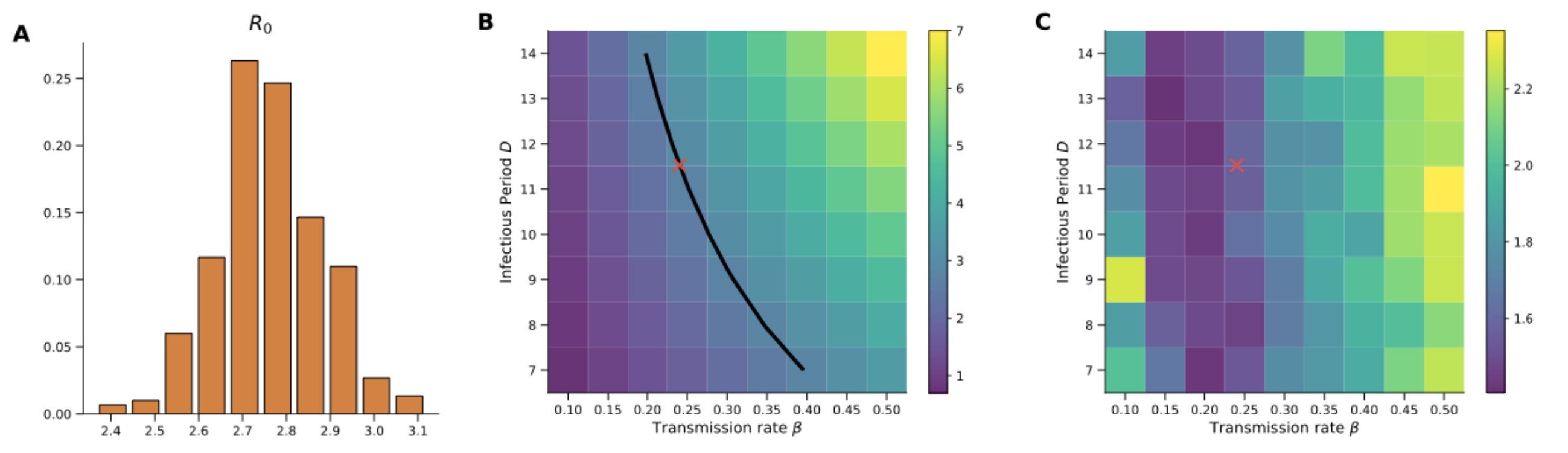}
    \caption{Sensitivity and identifiability analysis of estimated R0.}
    \label{fig2}
\end{figure}

To further evaluate the transmission potential of the Hantavirus variant in the cruise ship environment, we estimated the basic reproductive number, $R_0$, before the implementation of strict cabin quarantine measures. As shown in Figure~\ref{fig2}A, the estimated $R_0$ distribution has a mean of 2.76, with a 95\% confidence interval of 2.52--2.99. This elevated $R_0$ indicates substantial potential for onboard transmission prior to quarantine.

We further conducted sensitivity analyses to assess the identifiability of key epidemiological parameters within our model-inference framework. Specifically, we evaluated the effects of varying the transmission rate, $\beta$, and the infectious period, $D$, on the effective reproductive number and overall model fit, while holding all other parameters fixed at their mean estimates, as in Figure~\ref{fig1}. In Figure~\ref{fig2}B, the black solid line denotes $R_0 = 2.76$, and the red cross marks the mean estimates of $\beta$ and $D$ inferred by the model.

To determine the optimal parameter combination, we calculated the root mean square error (RMSE) between model simulations and ground-truth observations across different values of the transmission rate, $\beta$, and the infectious period, $D$. As shown in Figure~\ref{fig2}C, we discretized the parameter space into an $8 \times 9$ grid, performed simulations at each grid point, and computed the corresponding RMSE. The heat map shows that the lowest-RMSE region is strongly centered around the best-fit parameter values, $\beta \approx 0.24$ and $D \approx 11.52$, indicated by the red cross. The RMSE increases substantially as the parameter values move away from this optimum, supporting the identifiability of these parameters under the SEIRD model structure and the available daily incidence observations from the cruise ship outbreak.

\subsection{Missing exposed cases}

\begin{figure}[h]
    \centering
    \includegraphics[width=0.6\linewidth]{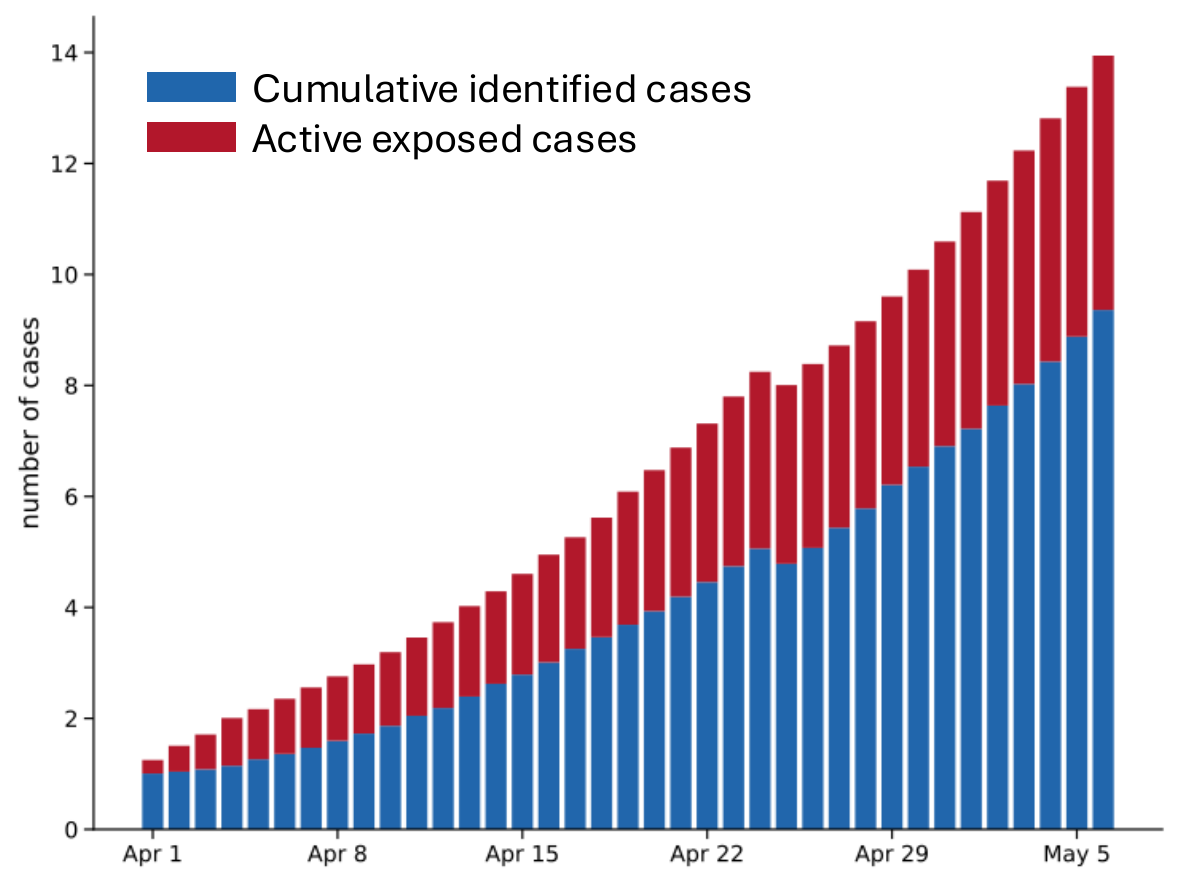}
    \caption{Comparison between identified cases and active exposed cases.}
    \label{fig3}
\end{figure}

We further compared the number of identified cases with the number of active exposed cases. Since exposed individuals may subsequently transition to identified cases, we focus here on active exposed cases, defined as individuals who remain in the exposed state and have not yet progressed to the symptomatic or otherwise identifiable stage. As shown in Figure~\ref{fig3}, several exposed cases remain unidentified, revealing a substantial reservoir of unisolated exposed passengers. This finding underscores that relying solely on symptom-based surveillance can severely underestimate the true scale of infection, highlighting the urgent need for active surveillance and widespread testing to identify and isolate exposed individuals before further transmission occurs.

\section{Discussion}

This study presents an initial analysis of the transmission dynamics of a Hantavirus variant within the distinctive enclosed environment of a cruise ship. By examining a recent outbreak aboard a commercial vessel, the study addresses a timely public health concern and provides practical insights into the management of acute infectious-disease events in high-density, confined settings.

Our results indicate that the basic reproduction number, \(R_0\), in the cruise ship environment is substantially greater than 1, suggesting that sustained onboard transmission may occur in the absence of stringent control measures. The simulation results further reveal a considerable population of individuals in the exposed compartment who have not yet been detected or isolated. Because these latent or pre-symptomatic infections may lack clear clinical manifestations, such individuals can remain unidentified and may contribute to continued transmission, thereby increasing the potential scale of the outbreak. Consequently, the timely identification, monitoring, and tracing of exposed individuals remains a critical and challenging component of effective epidemic control.

Despite its value for risk assessment, this study has several limitations. The current model relies on established biological characteristics and epidemiological assumptions derived from known Hantavirus strains. However, the infectiousness, latency period, and transmission pathways of this emerging variant remain incompletely characterized. As additional epidemiological data become available, particularly data describing the variant-specific natural history and transmission patterns, the proposed model-inference framework can be further refined, calibrated, and validated.

Overall, this study underscores the substantial transmission risk posed by this Hantavirus variant in enclosed and densely populated environments. The findings provide theoretical insight into outbreak dynamics and offer practical guidance for cruise operators and public health authorities. In particular, they may support the design of future epidemic-response strategies, the optimization of early-screening and surveillance protocols, and the implementation of measures to protect the health of passengers and crew.

\newpage

\bibliographystyle{ieeetr}
\bibliography{references}

@website{WHO,
    title = {https://www.who.int/news/item/07-05-2026-who-s-response-to-hantavirus-cases-linked-to-a-cruise-ship},
}

@website{ECDC,
    title = {https://www.ecdc.europa.eu/sites/default/files/documents/TAB-hantavirus-06052026.pdf},
}

@article{kak2007infections,
  title={Infections in confined spaces: cruise ships, military barracks, and college dormitories},
  author={Kak, Vivek},
  journal={Infectious disease clinics of North America},
  volume={21},
  number={3},
  pages={773--784},
  year={2007},
  publisher={Elsevier}
}

@article{zhang2016contact,
  title={Contact infection of infectious disease onboard a cruise ship},
  author={Zhang, Nan and Miao, Ruosong and Huang, Hong and Chan, Emily YY},
  journal={Scientific reports},
  volume={6},
  number={1},
  pages={38790},
  year={2016},
  publisher={Nature Publishing Group UK London}
}

@article{hu2013scaling,
  title={The scaling of contact rates with population density for the infectious disease models},
  author={Hu, Hao and Nigmatulina, Karima and Eckhoff, Philip},
  journal={Mathematical biosciences},
  volume={244},
  number={2},
  pages={125--134},
  year={2013},
  publisher={Elsevier}
}

@article{rocklov2020covid,
  title={COVID-19 outbreak on the Diamond Princess cruise ship: estimating the epidemic potential and effectiveness of public health countermeasures},
  author={Rockl{\"o}v, Joacim and Sj{\"o}din, Henrik and Wilder-Smith, Annelies},
  journal={Journal of travel medicine},
  volume={27},
  number={3},
  pages={taaa030},
  year={2020},
  publisher={Oxford University Press}
}

@article{emery2020contribution,
  title={The contribution of asymptomatic SARS-CoV-2 infections to transmission on the Diamond Princess cruise ship},
  author={Emery, Jon C and Russell, Timothy W and Liu, Yang and Hellewell, Joel and Pearson, Carl AB and Knight, Gwenan M and Eggo, Rosalind M and Kucharski, Adam J and Funk, Sebastian and others},
  journal={Elife},
  volume={9},
  pages={e58699},
  year={2020},
  publisher={eLife Sciences Publications, Ltd}
}

@article{batista2020minimizing,
  title={Minimizing disease spread on a quarantined cruise ship: A model of COVID-19 with asymptomatic infections},
  author={Batista, Berlinda and Dickenson, Drew and Gurski, Katharine and Kebe, Malick and Rankin, Naomi},
  journal={Mathematical biosciences},
  volume={329},
  pages={108442},
  year={2020},
  publisher={Elsevier}
}

@article{pavli2016respiratory,
  title={Respiratory infections and gastrointestinal illness on a cruise ship: a three-year prospective study},
  author={Pavli, Androula and Maltezou, Helena C and Papadakis, Antonis and Katerelos, Panagiotis and Saroglou, Georgios and Tsakris, Athanasios and Tsiodras, Sotirios},
  journal={Travel medicine and infectious disease},
  volume={14},
  number={4},
  pages={389--397},
  year={2016},
  publisher={Elsevier}
}

@article{manigold2014human,
  title={Human hantavirus infections: epidemiology, clinical features, pathogenesis and immunology},
  author={Manigold, Tobias and Vial, Pablo},
  year={2014},
  publisher={Swiss Society of Internal Medicine, Swiss Society of Infectiology and Swiss~…}
}

@article{young2000incubation,
  title={The incubation period of hantavirus pulmonary syndrome.},
  author={Young, Joni C and Hansen, Gail R and Graves, Tim K and Deasy, Marshall P and Humphreys, Jan G and Fritz, Curtis L and Gorham, Kassandra L and Khan, Ali S and Ksiazek, Thomas G and Metzger, Kristina B and others},
  journal={The American journal of tropical medicine and hygiene},
  volume={62},
  number={6},
  pages={714--717},
  year={2000}
}

@article{mizumoto2020estimating,
  title={Estimating the asymptomatic proportion of coronavirus disease 2019 (COVID-19) cases on board the Diamond Princess cruise ship, Yokohama, Japan, 2020},
  author={Mizumoto, Kenji and Kagaya, Katsushi and Zarebski, Alexander and Chowell, Gerardo},
  journal={Eurosurveillance},
  volume={25},
  number={10},
  pages={2000180},
  year={2020}
}

@article{zhang2020estimation,
  title={Estimation of the reproductive number of novel coronavirus (COVID-19) and the probable outbreak size on the Diamond Princess cruise ship: A data-driven analysis},
  author={Zhang, Sheng and Diao, MengYuan and Yu, Wenbo and Pei, Lei and Lin, Zhaofen and Chen, Dechang},
  journal={International journal of infectious diseases},
  volume={93},
  pages={201--204},
  year={2020},
  publisher={Elsevier}
}

@article{vial2006incubation,
  title={Incubation period of hantavirus cardiopulmonary syndrome},
  author={Vial, Pablo A and Valdivieso, Francisca and Mertz, Gregory and Castillo, Constanza and Belmar, Edith and Delgado, Iris and Tapia, Mauricio and Ferr{\'e}s, Marcela},
  journal={Emerging Infectious Diseases},
  volume={12},
  number={8},
  pages={1271},
  year={2006}
}

@article{hethcote2000mathematics,
  title={The mathematics of infectious diseases},
  author={Hethcote, Herbert W},
  journal={SIAM review},
  volume={42},
  number={4},
  pages={599--653},
  year={2000},
  publisher={SIAM}
}

@article{diekmann1990definition,
  title={On the definition and the computation of the basic reproduction ratio R 0 in models for infectious diseases in heterogeneous populations},
  author={Diekmann, Odo and Heesterbeek, Johan Andre Peter and Metz, Johan Anton Jacob},
  journal={Journal of mathematical biology},
  volume={28},
  number={4},
  pages={365--382},
  year={1990},
  publisher={Springer}
}

@article{van2002reproduction,
  title={Reproduction numbers and sub-threshold endemic equilibria for compartmental models of disease transmission},
  author={Van den Driessche, Pauline and Watmough, James},
  journal={Mathematical biosciences},
  volume={180},
  number={1-2},
  pages={29--48},
  year={2002},
  publisher={Elsevier}
}

@article{chopra2022differentiable,
  title={Differentiable agent-based epidemiology},
  author={Chopra, Ayush and Rodr{\'\i}guez, Alexander and Subramanian, Jayakumar and Quera-Bofarull, Arnau and Krishnamurthy, Balaji and Prakash, B Aditya and Raskar, Ramesh},
  journal={arXiv preprint arXiv:2207.09714},
  year={2022}
}

@article{martinez2020super,
  title={“Super-spreaders” and person-to-person transmission of Andes virus in Argentina},
  author={Mart{\'\i}nez, Valeria P and Di Paola, Nicholas and Alonso, Daniel O and P{\'e}rez-Sautu, Unai and Bellomo, Carla M and Iglesias, Ayel{\'e}n A and Coelho, Rocio M and L{\'o}pez, Beatriz and Periolo, Natalia and Larson, Peter A and others},
  journal={New England Journal of Medicine},
  volume={383},
  number={23},
  pages={2230--2241},
  year={2020},
  publisher={Mass Medical Soc}
}

@article{lazaro2007clusters,
  title={Clusters of hantavirus infection, southern Argentina},
  author={L{\'a}zaro, Maria E and Cantoni, Gustavo E and Calanni, Liliana M and Resa, Amanda J and Herrero, Eduardo R and Iacono, Marisa A and Enria, Delia A and Cappa, Stella M Gonz{\'a}lez},
  journal={Emerging Infectious Diseases},
  volume={13},
  number={1},
  pages={104},
  year={2007}
}

@article{li2020substantial,
  title={Substantial undocumented infection facilitates the rapid dissemination of novel coronavirus (SARS-CoV-2)},
  author={Li, Ruiyun and Pei, Sen and Chen, Bin and Song, Yimeng and Zhang, Tao and Yang, Wan and Shaman, Jeffrey},
  journal={Science},
  volume={368},
  number={6490},
  pages={489--493},
  year={2020},
  publisher={American Association for the Advancement of Science}
}

@article{pei2020differential,
  title={Differential effects of intervention timing on COVID-19 spread in the United States},
  author={Pei, Sen and Kandula, Sasikiran and Shaman, Jeffrey},
  journal={Science advances},
  volume={6},
  number={49},
  pages={eabd6370},
  year={2020},
  publisher={American Association for the Advancement of Science}
}

\end{document}